\documentstyle[twocolumn,aps,epsf]{revtex}

\def\i{{\rm i}}
\def\d{{\rm d}}
\def\e{{\rm e}}
\def\vector#1{{\bf #1}}
\def\vp{{\vector p}}

\def\vq{{\vector q}}

\def\vr{{\vector r}}

\def\vB{{\vector B}}
\def\vvF{{\vector v}_{\rm F}}
\def\vphat{{\hat {\vector p}}}

\def\Tc{{T_{\rm c}}}

\def\hightc{{high-$T_{\rm c}$ }}

\def\RSGCO{{${\rm RuSr_2GdCu_2O_8}$}}
\def\YBCO{{${\rm YBa_2Cu_3O_7}$}}
\def\YBCOdelta{{${\rm YBa_2Cu_3O_{7+\delta}}$}}

\def\RCeRSCO{{${\rm {\it R}_{1.4}Ce_{0.6}RuSr_2Cu_2O_{10-\delta}}$}}

\def\hsp#1{\hspace{#1ex}}

\def\Tc{{T_{\rm c}}}

\def\lsim{\stackrel{{\textstyle<}}{\raisebox{-.75ex}{$\sim$}}}
\def\gsim{\stackrel{{\textstyle>}}{\raisebox{-.75ex}{$\sim$}}}

\def\ppara{{p_{\parallel}}}
\def\hc{{h_{\rm c}}}
\def\HP{{H_{\rm P}}}
\def\omegaD{{\omega_{\rm D}}}
\def\lsub#1{{\stackrel{\mbox{}}{\raisebox{-1.5ex}{\scriptsize $#1$}}}}

\begin{document}
\draft

\twocolumn[\hsize\textwidth\columnwidth\hsize\csname 
@twocolumnfalse\endcsname

\title{
Superconductivity in a Ferromagnetic Layered Compound
} 

\author{Hiroshi Shimahara and Satomi Hata} 

\def\runtitle{Superconductivity in a Ferromagnetic Layered Compound}


\address{
Department of Quantum Matter Science, ADSM, Hiroshima University, 
Higashi-Hiroshima 739-8526, Japan
}

\date{Received ~~~ January 2000}

\maketitle

\begin{abstract}
We examine superconductivity in layered systems with large Fermi-surface 
splitting due to coexisting ferromagnetic layers. 
In particular, the hybrid ruthenate-cuprate compound \RSGCO \hsp{0.25} 
is examined on the coexistence of the superconductivity and the 
ferromagnetism, which has been observed recently. 
We calculate critical fields of the superconductivity 
taking into account the Fulde-Ferrell-Larkin-Ovchinnikov state 
in a model with Fermi-surfaces which shapes are similar to 
those obtained by a band calculation. 
It is shown that the critical field is enhanced remarkably 
due to a Fermi-surface effect, 
and can be high enough to make the coexistence possible 
in a microscopic scale. 
We also clarify the direction of the spatial oscillation of the order 
parameter, which may be observed by scanning tunneling microscope 
experiments. 
\end{abstract}

\pacs{
}


]

\narrowtext

Recently, coexistence of superconductivity and ferromagnetism 
has been reported in the hybrid ruthenate-cuprate 
compounds 
\RCeRSCO \hsp{0.25} ($R = {\rm Eu}$ and Gd) 
and 
\RSGCO~\cite{Ber99,Pri99,Fel99}. 
These compounds have similar crystal structures to the \hightc cuprate 
superconductor \YBCO \hsp{0.5} except that layers of ${\rm CuO}$ chains 
are replaced with ruthenate layers. 
Experimental and theoretical studies indicate that the ruthenate layers 
are responsible for the ferromagnetic long range order~\cite{Fel99,Pic99}, 
while the cuprate layers for the superconductivity~\cite{Fel99}.

One of the remarkable features of these compounds is that 
the superconducting transition occurs at a temperature well below the 
ferromagnetic transition temperature unlike most of the other ferromagnetic 
superconductors. 
For example, in \RSGCO, 
the superconducting transition was observed at $T_{\rm c} \sim 46{\rm K}$, 
whereas the ferromagnetic transition 
at $T_{\rm M} \sim 132{\rm K}$~\cite{Ber99}. 
Therefore, the ferromagnetic order can be regarded as a rigid back 
ground which is not modified very much by the appearance of 
the superconductivity. 
This picture is also supported by experimental 
observations~\cite{Ber99,Pri99,Fel99}.

According to the first principle calculations 
by Pickett {\it et al.}~\cite{Pic99}, 
magnetic fields in the cuprate layers due to the ordered spin moment 
in the ruthenate layers 
are much smaller than exchange fields mediated by electrons. 
The exchange fields play a role like magnetic fields which act 
only on the spin digrees of freedom but do not create Lorentz force. 
Therefore, the present system is approximately equivalent 
to a quasi-two-dimensional system in magnetic fields nearly 
parallel to the layers.

However, such Fermi-surface splitting gives rise to pair-breaking effect 
as well as that due to a parallel magnetic field. 
The exchange field in \RSGCO \hsp{0.25} is very large and seems to 
exceed the Pauli paramagnetic limit 
(Chandrasekar-Clogston limit)~\cite{NotePl}. 
The Pauli paramagnetic limit $\HP$ at $T = 0$ is roughly estimated 
from the zero field transition temperature $T_{\rm c}^{(0)}$ 
by a simplified formula 
$\mu_{e} \HP = 1.25 \hsp{0.3} T_{\rm c}^{(0)}$, 
where $\mu_{e}$ denotes the electron magnetic moment. 
For \RSGCO, since the exchange field exists in practice, 
$T_{\rm c}^{(0)}$ of isolated cuprate layers 
is not known, but it will be appropriate to assume 
$T_{\rm c}^{(0)} \lsim 90{\rm K}$ 
from the transition temperature of \YBCOdelta \hsp{0.25} 
at the optimum electron density. 
Hence we obtain $\mu_{e} \HP \lsim 110 {\rm K}$ at $T = 0$ 
from the above formula. 
On the other hand, the band calculation gives an estimation 
$\mu_{e} B_{\rm ex} = \Delta_{\rm ex}/2 
\sim 25{\rm meV}/2 \sim 107{\rm K}$~\cite{Pic99}. 
It is remarkable that the superconducting transition occurs at 
such a high temperature $\Tc \approx 46{\rm K}$ 
in spite of the strong exchange field of the order of 
the Pauli paramagnetic limit at $T = 0$.

There are some mechanisms by which the critical field 
of superconductivity exceeds the Pauli limit. 
For example, the triplet pairing superconductivity is 
an important candidate. 
However, from their crystal structures and high transition temperatures, 
it is plausible that the present compounds are categorized as \hightc 
cuprate superconductors and 
therefore the superconductivity is due to an anisotropic singlet pairing 
with line nodes, which is conventionally called a $d$-wave pairing. 
For the singlet pairing, possibility of 
an inhomogeneous superconducting state that is called 
a Fulde-Ferrell-Larkin-Ovchinnikov (FFLO or LOFF) 
state~\cite{Ful64,Lar64} was discussed 
by Pickett {\it et al.}~\cite{Pic99} 
as a candidate for the mechanism.

On the possibility of the FFLO state, they pointed out 
that there are nearly flat areas in the Fermi-surfaces in \RSGCO, 
which favor the FFLO state. 
It is known that the FFLO critical field diverges at $T=0$ 
in one dimensional models. 
However, if the Fermi-surfaces are too flat, 
nesting instabilities, such as those to spin density wave (SDW) and 
charge density wave (CDW), are favored for realistic interaction 
strengths. 
For the present compound, the nearly flat areas are not 
so flat that the nesting instabilities occur, 
but the small curvature still enhances the FFLO state~\cite{Shi94}.

It is also known that even in the absence of the flat areas, 
the critical field is enhanced in the two-dimensional (2D) systems 
in comparison to 
the three dimensional systems~\cite{Bur94,Shi94,Dup95,Shi97a}. 
Further, when the Fermi-surface structure of the system satisfies 
a certain condition, the critical field can reach several times 
the Pauli limit even in the absence of nearly flat areas~\cite{Shi99}. 
Such a Fermi-surface effect can be regarded as a kind of nesting effects 
analogous to those for SDW and CDW~\cite{Shi94}. 
The ``nesting'' effect was examined in details in our previous papers, 
where 2D tight binding models are studied as examples~\cite{Shi97a,Shi99}.

Direct evidence of the FFLO state may be obtained 
by scanning tunneling microscope (STM) experiments. 
For a comparison with experimental results, spatial structure of 
the order parameter should be predicted theoretically. 
In particular, direction of the modulation of the order parameter is 
important. 
It may appear that the modulation must be in the direction 
perpendicular to the flattest area of the Fermi-surface, 
because then the spatial variation is minimized. 
However, in some of 2D models, it is not perpendicular 
to flattest areas~\cite{Shi97a,Shi99}. 
Only explicite calculations which take into account the Fermi-surface 
structure could clarify the direction of the modulation.

Therefore, the purposes of this paper are 
(1) estimation of the critical field of superconductivity including 
the FFLO state to examine the possibility of coexistence of singlet 
superconductivity and ferromagnetism in a microscopic scale, 
and (2) clarification of the direction of the spatial oscillation 
of the order parameter to compare with results of STM experiments 
possible in the future. 
We examine a tight binding model with Fermi-surfaces which shapes 
are similar to those of \RSGCO, 
because the quantities that we are calculating are sensitive to 
the Fermi-surface structure.

Recently, the FFLO state has been studied in a tight binding model 
with only nearest neighbor hopping~\cite{Shi99,Zhu00}. 
It was found that ratio of the FFLO critical field and the Pauli limit 
is small near the half filling. 
Zhu {\it et al.} have discussed that hence the coexistence of the 
superconductivity and the ferromagnetic order is difficult 
except in the vicinity of the ferromagnetic domains 
near the half filling~\cite{Zhu00,Chu99}. 
However, some experimental results indicate coexistence 
in a microscopic scale and a bulk Meissner-state~\cite{Ber99,Ber00}. 
Here, we should note that the tight binding model with only nearest 
neighbor hopping can not reproduce the shapes of the Fermi-surfaces 
of \RSGCO. 
By taking into account the realistic Fermi-surface structure, 
we will show below that the critical field is enhanced remarkably 
and thus the coexistence in a microscopic scale is possible 
in this compound.

First, we define the tight binding model 
\def\eqdefmodel{(1)}
$$
     H_0 = \sum_{\vp \sigma} \epsilon_{\vp \sigma} 
     c_{\vp \sigma}^{\dagger}
     c_{\vp \sigma}
     \eqno\eqdefmodel
     $$
with a dispersion relation 
\def\eqdefeps{(2)}
$$
     \epsilon_{\vp \sigma} = 
     - 2 t ( \cos p_x + \cos p_y )
     - 4 t_2 \cos p_x \cos p_y 
     - \mu + h \sigma , 
     \eqno\eqdefeps
     $$
where $h$ denotes the exchange field. 
When we apply the present theory to type II superconductors 
in a magnetic field $\vB$, $h$ is written as $h = \mu_{e} |\vB|$. 
We use a unit with $t = 1$ and the lattice constant $a = 1$ 
in this paper.

\begin{figure}[htb]
\begin{center}
\leavevmode \epsfxsize=7cm  
\epsfbox{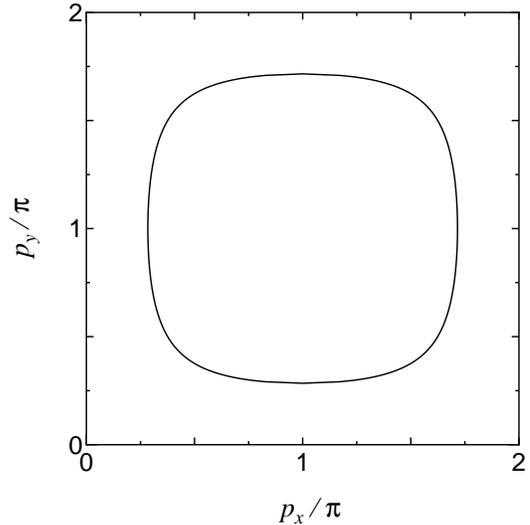}
\end{center}
\caption{Fermi-surface of the present model Hamiltonian for 
$t_2 = -0.6$ and $n = 1.1$. } 
\label{fig:FS}
\end{figure}

We take the value of the second nearest neighbor hopping energy 
$t_2 = - 0.6 t$, 
which gives shapes of the Fermi-surfaces similar to 
the symmetric ${\rm CuO_2}$ barrel Fermi-surfaces obtained 
by Pickett {\it et al.}~\cite{Pic99} 
at $n = 1.1$ as shown in Fig.\ref{fig:FS}. 
Here, $n$ is the electron number par a site.

We calculate the critical field in the ground state 
for $n= 0.92 \sim 2$, 
applying a formula developed in our previous papers~\cite{Shi97a,Shi99}. 
For anisotropic pairing 
\def\eqDeltaalpha{(3)}
$$
     \Delta(\vphat,\vr) = \Delta_{\alpha} \gamma_{\alpha}(\vphat) 
                          \, \e^{\i \vq \cdot \vr}  
     \eqno\eqDeltaalpha
     $$
($\vphat \equiv \vp/|\vp|$), 
the critical field is give by 
\def\eqhc{(4)}
$$
     h_{\rm c} = \max_{\scriptsize \vq} {\Bigl [}
     \frac{\Delta_{\alpha 0}}{2} \exp {\bigl (} 
          - \int \frac{\d \ppara}{2\pi}
            \frac{\rho_{\perp}^{\alpha}(0,\ppara)}
                 {N_{\alpha}(0)}
            \log |1- \frac{{\vvF} \cdot {\vq}}{2\hc}| 
     {\bigr )} {\Bigr ]} , 
     \eqno\eqhc
     $$
where 
$\Delta_{\alpha 0} \equiv 
2\omegaD \exp(-1/g_{\alpha}N_{\alpha}(0))\approx 1.76k_{\rm B}\Tc$ 
and $\rho_{\perp}^{\alpha}(0,\ppara)
        \equiv \rho_{\perp}(0,\ppara) [\gamma_{\alpha}(\vphat)]^2$ 
with the momentum dependent density of states 
$\rho_{\perp}(\epsilon,\ppara)$. 
Here, $\ppara$ denotes the momentum component along the Fermi-surface. 
The pairing interaction is assumed to have a form 
\def\eqVdef{(5)}
$$
     V(\vp,\vp') = - g_\alpha \gamma_\alpha (\vphat) 
                              \gamma_\alpha (\vphat') . 
     \eqno\eqVdef
     $$
In particular, for $d$-wave pairing, we use a model with 
\def\eqgammad{(6)}
$$
     \gamma_d(\vphat) \propto \cos p_x - \cos p_y , 
     \eqno\eqgammad
     $$
where $p_x$ and $p_y$ are the momentum components on the Fermi-surface 
in the directions of $\vphat$. 
In our previous papers, it was shown that the qualitative and 
semi-quantitative results are not sensitive to details of the form 
of $\gamma_d(\vphat)$~\cite{Shi97a,Shi99}. 
An effective density of states $N_{\alpha}(0)$ for anisotropic pairing 
is defined by 
\def\eqNalphazero{(7)}
$$
     N_{\alpha}(0) \equiv N(0) \langle [\gamma_{\alpha}(\vphat)]^2 \rangle , 
     \eqno\eqNalphazero
     $$
with an average on the Fermi-surface 
\def\eqdefave{(8)}
$$
     \langle \cdots \rangle 
          = \int \frac{\d \ppara}{2\pi}
                 \frac{\rho_{\perp}(0,\ppara)}{N(0)}
                 (\cdots) \lsub {|\vp|=p_{\rm F}(\ppara)} , 
     \eqno\eqdefave
     $$
where $N(0)$ is the density of states at the Fermi level. 
The Pauli limit $\HP$ for anisotropic pairing is calculated by 
\def\eqHP{(9)}
$$
     \mu_{e} \HP 
        = \frac{\sqrt{ \langle [\gamma_{\alpha}(\vphat)]^2 \rangle}}
               {{\bar \gamma_{\alpha}}}
          \frac{\Delta_{\alpha 0}}{\sqrt{2}} 
     \eqno\eqHP
     $$
with 
\def\eqgammabar{(10)}
$$
     \frac{1}{{\bar \gamma_{\alpha}}} = 
     \exp{\bigl (}
     \frac{ \langle [\gamma_{\alpha}(\vphat)]^2 
                    \log[1/|\gamma_{\alpha}(\vphat)|] \rangle}
          { \langle [\gamma_{\alpha}(\vphat)]^2 \rangle}
     {\bigr )} . 
     \eqno\eqgammabar
     $$

In the above equations, the vector $\vq$ is the center-of-mass momentum 
of Cooper pairs of the FFLO state. 
From the symmetry of the system, there are four or eight 
equivalent optimum vectors ($\vq_m$'s), 
depending on whether $\vq$ is in a symmetry direction or not, 
respectively. 
Actually, arbitrary linear combination of $\exp(\i \vq_m \cdot \vr)$ 
gives the same second order critical field, 
and the degeneracy is removed by the nonlinear term of the gap 
equation below the critical field~\cite{Lar64,Shi98}. 
However, regarding the critical field and the optimum direction 
of the oscillation of the order parameter near the critical field, 
it is sufficient to take a single $\vq$ as in eq.{\eqDeltaalpha}.

\begin{figure}[htb]
\begin{center}
\leavevmode \epsfxsize=7cm  
\epsfbox{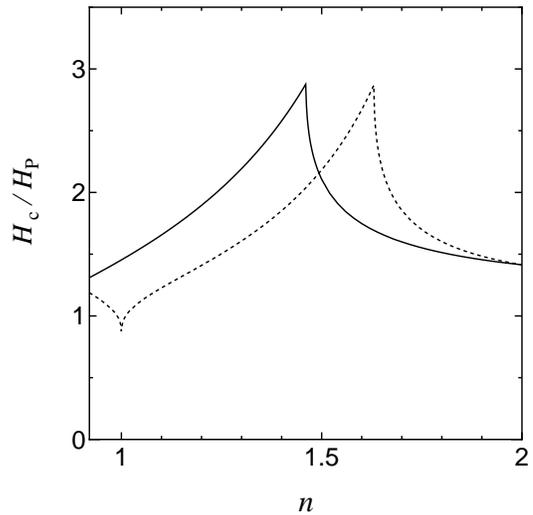}
\end{center}
\caption{
Critical fields of the FFLO state of the $s$-wave pairing 
for $n = 0.92 \sim 2$ at $T = 0$. 
Solid and broken lines show the results for $t_2=-0.6$ 
and $t_2 = 0$, respectively.} 
\label{fig:hcs}
\end{figure}

\begin{figure}[htb]
\begin{center}
\leavevmode \epsfxsize=7cm  
\epsfbox{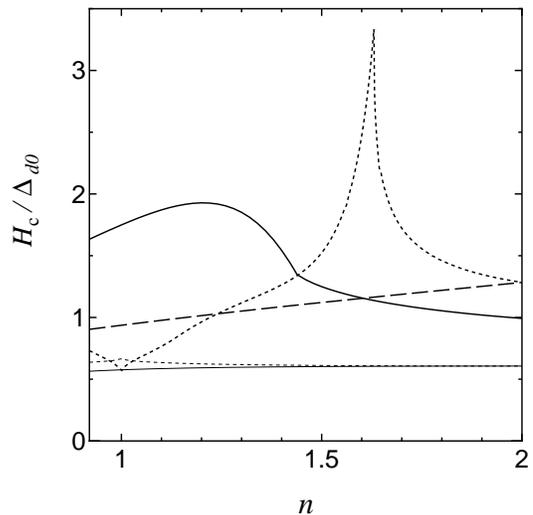}
\end{center}
\caption{
Critical fields of the FFLO state of the $d$-wave pairing 
for $n = 0.92 \sim 2$ at $T = 0$. 
Solid and broken lines show the results for 
$\varphi_{\vq}=\pi/4$ and $\varphi_{\vq}=0$, respectively, 
when $t_2=-0.6$. 
The dotted line shows the result for $t_2 = 0$. 
Thin solid line and thin dotted line show the Pauli paramagnetic 
limits in the unit of $\Delta_{d0}$, 
for $t_2 = -0.6$ and $0$, respectively.}
\label{fig:hcd}
\end{figure}

Figures \ref{fig:hcs} and \ref{fig:hcd} show numerical results of 
the critical fields for $t_2 = -0.6$, 
with our previous results for $t_2 = 0$ (dotted lines)~\cite{Shi99}. 
It is found that the critical fields are remarkably enhanced near 
the electron densities $n \approx 1.46$ and $1.20$ 
for the $s$-wave and the $d$-wave pairing, respectively. 
For example, at the electron density $n = 1.1$, 
the ratios of the critical field to the Pauli paramagnetic limit 
are approximately equal to 1.66 and 3.19 
for the $s$-wave and the $d$-wave pairing, respectively. 
These values (especially the latter) seem to be 
large enough to make the coexistence possible in \RSGCO.

In Fig.\ref{fig:hcd} for the $d$-wave pairing, 
both the critical fields for $\varphi_{\vq} = \pi/4$ and 
$\varphi_{\vq} = 0$ are shown, 
but the highest one is the final result of the critical field 
given by eq.{\eqhc}. 
Here, $\varphi_{\vq}$ is the angle 
between the optimum $\vq$ and one of the crystal axes. 
It is shown by a numerical calculation that the critical fields for 
the other values of $\varphi_{\vq}$ are lower than the higher one of 
the critical fields for $\varphi_{\vq} = \pi/4$ and 0. 
Thus, the direction of the optimum wave vector $\vq$ jumps from 
$\varphi_{\vq} = \pi/4$ to $\varphi_{\vq} = 0$ at $n \approx 1.63$. 
On the other hand, for the $s$-wave pairing, $\varphi_{\vq} = \pi/4$ 
is the optimum in the whole region of the electron density. 
These behaviors are different from that for $t_2 = 0$, in which 
$\varphi_{\vq} = 0$~\cite{Shi99}.

For $t_2 = -0.6$, a cusp is seen in Fig.\ref{fig:hcs} 
for the $s$-wave pairing, 
whereas it does not appear in Fig.\ref{fig:hcd} 
for the $d$-wave pairing. 
The physical origin of the cusp at $n \approx 1.46$ is that 
the Fermi surfaces satisfy a certain condition there, 
which was explained in our previous paper for $t_2 = 0$~\cite{Shi99}. 
It is related to how the two Fermi-surfaces touch 
by the translation by the optimum $\vq$. 
In the present case ($t_2 = -0.6$ and $n \approx 1.46$), 
the touch occurs in the (110) direction, 
but because of the nodes of the order parameter 
the ``nesting'' is not efficient for the $d$-wave pairing. 
Therefore, cusp does not appear for the $d$-wave pairing.

\begin{figure}[htb]
\begin{center}
\leavevmode \epsfxsize=7cm  
\epsfbox{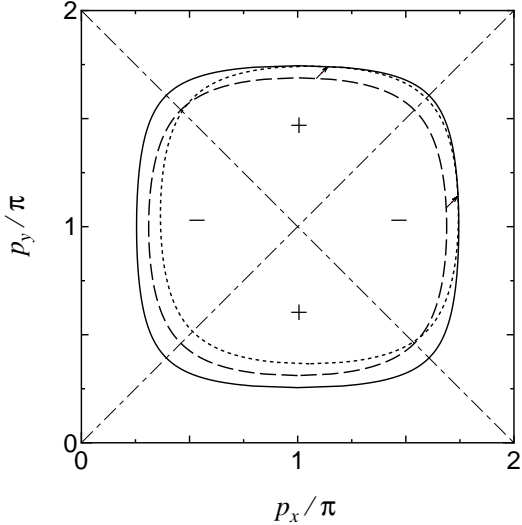}
\end{center}
\caption{
Fermi-surface nesting for the FFLO state and the optimum 
wave vector $\vq$ of the FFLO state at the critical field 
($h_{\rm c} = 1.87 \Delta_{d0}$) at $n = 1.1$. 
Solid and broken lines show the Fermi-surfaces of the up and down spin 
electrons, respectively. 
Dotted line shows the Fermi-surface of the down spin electrons shifted 
by $\vq$. 
Small arrows show the wave vector $\vq$. 
They are placed at the momenta 
at which the two Fermi-surfaces touch. 
Dotted broken lines show the nodes of the $d$-wave order parameter. 
We use a large value of $\Delta_{d0} = 0.3t/1.87$ 
({\it i.e.}, $h_{\rm c} = 0.3t$) 
in order to make the displacement visible. }
\label{fig:Nes}
\end{figure}

In spite of the absence of cusp behavior, the critical field is 
still very large for the $d$-wave pairing near the half-filling. 
Figure \ref{fig:Nes} shows the nesting behavior of the Fermi-surfaces 
at $t_2 = -0.6$ and $n = 1.1$. 
The direction of the optimum vector $\vq$ is $\varphi_{\vq} = \pi/4$, 
and the Fermi-surfaces touch at two points 
({\it i.e.}, two lines in the $p_x p_y p_z$-space), 
$(p_x,p_y) \approx (1.113 \pi, 1.713 \pi)$ and $(1.713 \pi, 1.113 \pi)$. 
Since $\varphi_{\vq} = \pi/4$ is also the direction of a node 
of the $d$-wave order parameter, 
it may appear that this direction is less favorable. 
However, in actuality the critical field is remarkably enhanced 
for this ``nesting'' vector $\vq$, 
since it gives two nesting lines which are far away from the nodes 
but near the flattest areas, as shown in Fig.\ref{fig:Nes}. 
Besides, they are near both the maxima of the $d$-wave order parameter 
and the van Hove singularities, 
which also enhance the critical field. 
As the electron density increases, the two nesting lines 
approach to the line node of the order parameter, 
and thus the critical field decreases.

Since the optimum direction $\varphi_{\vq} = \pi/4$ is 
in a symmetry line, 
there are four equivalent directions, that is, 
$\varphi_{\vq} = \pm \pi/4$ and $\pm 3\pi/4$. 
Therefore, symmetric linear combinations such as 
\def\eqDstruc{(11)}
$$
     \begin{array}{rcl} 
     \Delta(\vp,\vr) & \propto & 
       \cos(q x') \\
     \Delta(\vp,\vr) & \propto & 
       \cos(q x') + \cos(q y') \\
     \end{array}
     \eqno\eqDstruc
     $$
are convincing candidates, which may be observed in the present 
compound, where $x' = (x+y)/\sqrt{2}$ and $y' = (x-y)/\sqrt{2}$. 
In particular, the 2D structures such as the latter of eq.{\eqDstruc} 
are favored at high fields~\cite{Shi98}.

For the FFLO state to appear, temperature needs to be lower 
than the tri-critical temperature $T^{*}$ 
of the FFLO, BCS and normal states. 
$T^{*}$ is generally equal to about $0.56 T_{\rm c}^{(0)}$ 
in simplified models such as eq.{\eqVdef}. 
If we apply this to the present system \RSGCO, 
$T^{*} \gsim T_{\rm c} \approx 46{\rm K}$ requires 
$T_{\rm c}^{(0)} \gsim 82{\rm K}$. 
This condition for $T_{\rm c}^{(0)}$ may be relaxed by taking into 
account a mixing of order parameters of different symmetries, 
which increases $T^{*}$~\cite{Mat94}.

In conclusion, the FFLO critical field of the cuprate layers is 
remarkably enhanced by an effect of the Fermi-surface structure. 
The direction of the spatial oscillation of the order parameter is 
in the $(110)$ direction both for the $s$-wave pairing 
and the $d$-wave pairing. 
Although we examined only the ground state in this paper, 
the result $H_{\rm c}/\HP \approx 3.19$ at $T = 0$ is large enough 
to support coexistence of the superconductivity and the ferromagnetic 
order in a microscopic scale in \RSGCO. 
Calculation for finite temperatures is now in progress. 

This work was supported by a grant for Core Research for Evolutionary 
Science and Technology (CREST) from Japan Science and Technology 
Corporation (JST).



\begin{thebibliography}{99}
\bibitem{NotePl}
  In this paper we use the term ``Pauli paramagnetic limit'' 
  for the BCS state with zero center-of-mass momentum.
\end{thebibliography}
\end{document}